\documentclass{PoS}
\usepackage{graphicx}

\newcommand{\newc}{\newcommand}
\newc{\gev}{\,GeV}
\newc{\sgn}{\mr{sgn}\,}
\newc{\ra}{\rightarrow}
\newc{\rpv}{$\mathrm{\not\!R_p}$}
\newc{\met}{$\not\!\!E_T$}
\newc{\rp}{$\mathrm{R_p}$}
\newc{\real}{\mathcal{R}e}
\newc{\alsm}{{\displaystyle \sum_{\alpha=1,2}}}
\newc{\besm}{{\displaystyle \sum_{\beta=1,2}}}
\newc{\al}{\alpha}
\newcommand{\ep}{\mbox{$\epsilon$}}
\catcode`@=11 
\def \gsim{\mathrel{\mathpalette\@versim>}}
\def \lsim{\mathrel{\mathpalette\@versim<}}
\def \@versim#1#2{\lower0.4ex\vbox{\baselineskip\z@skip\lineskip\z@skip
     \lineskiplimit\z@\ialign{$\m@th#1\hfil##\hfil$%
     \crcr#2\crcr\sim\crcr}}}
\catcode`@=12 
\def\gev{\: \rm GeV}

\title{${\cal {O}}(\alpha_s^2)$ QCD corrections to the resonant \\ sneutrino / slepton production at LHC }

\ShortTitle{NNLO QCD corrections}

\author{\speaker{Swapan K Majhi}\\
        Indian Association for the Cultivation of Science,\\
        2A\&2B Raja S C Mullick Road, Kolkata 700032, India\\
        E-mail: \email{swapan.majhi@saha.ac.in}}

\author{Prakash Mathews\\
        Saha Institute of Nuclear Physics,\\ 1/AF Bidhannagar, Kolkata 700064, India\\
        E-mail: \email{prakash.mathews@saha.ac.in}}

\author{V Ravindran\\
Regional Centre for Accelerator based Particle Physics,\\
Harish-Chandra Research Institute, Chhatnag Road, Jhusi,\\ 
Allahabad 211 019, India.\\
E-mail: \email{ravindra@hri.res.in}}

\abstract{We present a complete next to next to leading order QCD corrections to
the resonant production of sneutrino and charged slepton 
at the Tevatron and the Large Hadron Collider within the context of 
R-parity violating supersymmetric model.   We have demonstrated the role of
NNLO QCD corrections in reducing uncertainties resulting from 
renormalisation and factorisation scales and thereby making our predictions reliable.}

\FullConference{ 10th International Symposium on Radiative Corrections (Applications of Quantum Field Theory to Phenomenology) - Radcor2011\\
September 26-30, 2011\\
Mamallapuram, India}

\begin{document}

\section{Introduction}
In supersymmetry\cite{SUSY}, $R$-parity violation is one of the possible scenario in beyond standard model physics. With many interest, we consider only on the $\lambda'_{ijk}$ couplings which arises in lepton number violating term in  $R$-parity violation superpotential given below
\begin{equation}
    {\cal W} = \mu_i L_i H_2
                        + \lambda_{ijk} L_i L_j E^c_k
                        +  \lambda'_{ijk} L_i Q_j D^c_k
                        +  \lambda''_{ijk} U^c_i D^c_j D^c_k \ ,
      \label{eq:superpot}
\end{equation}
where $L_i$ and $Q_i $ are the $SU(2)$-doublet lepton and quark
superfields, $E^c_i, U^c_i, D^c_i$
the singlet superfields and $H_i$ the Higgs superfields.
The subscripts $i,j,k$ are generational indices. 
Note that $\lambda_{ijk}$ is antisymmetric
under the interchange of the first two indices and $\lambda''_{ijk}$
is antisymmetric under the interchange of the last two.
The first three terms in eqn.(\ref{eq:superpot}) violate
lepton number ($L$) and the last term violates baryon number ($B$) conservation.

Recently ATLAS group have been studied resonant production of heavy neutral scalar like sneutrino and subsequent decay to $e \mu$ final state. In their analysis, they put the bounds on sneutrino masses (see Ref.\cite{ATLAS}) on the basis of leading order (LO) result.
 In tevatron, both CDF\cite{CDF_col} and 
D0\cite{D0_col} collaboration 
analyse their data (Run-I as well as Run-II data) using our first results\cite{swapan} on the next to leading order (NLO) 
QCD corrections to sneutrino and charged slepton productions at hadron colliders. In their analysis to set bound on these R-parity violating couplings, 
cross section for
SM background processes namely Drell-Yan production of 
pair of leptons (say $l^+ l^- , l^\pm \nu$) 
(see first two papers of \cite{nnlody})
was considered at the next to next to leading order (NNLO) level while for the R-parity violating effects only NLO
corrected cross section was used. 
It was found that
the NLO QCD effects were quite large $\sim 10\% - 40\%$ at both Tevatron as well as 
LHC
Therefore, it is desirable to compute the cross sections 
for the resonant sneutrino and/or charged slepton productions at NNLO in QCD. 
These results will quantitatively improve the analysis
based on high statistics data available in the ongoing and future
experiments.  From the theoretical point of view, higher order 
radiative corrections provide a test of the convergence of the perturbation theory and
hence the reliable comparison of data with the theory predictions is possible. 
The fixed order perturbative results most often suffer from large uncertainties 
due to the presence of renormalisation and factorisation scales. 
They get reduced as we include more and more terms
in the perturbative expansion thanks to renormalisation group
invariance.  
In this article we have
systematically included its scale dependence through the renormalisation group equations 
and we discussed the impact of it in the next sections. 

\section{Brief discussion of NNLO calculations}
In this section, we describe very briefly, the computation of second order ($\alpha_s^2$) 
QCD radiative corrections to resonant production, in hadron colliders, 
of a sneutrino/charged slepton.
We present our results in such a way that 
they can be used for any scalar-pseudoscalar 
 production which
is the main goal of this work.  
The inclusive hadronic cross section for the reaction
\begin{eqnarray}
\label{eqn2.10}
H_1(P_1)+H_2(P_2)\rightarrow \phi (p_\phi)+X\,,
\end{eqnarray}
 is given by
\begin{eqnarray}
\label{eqn2.11}
&&\sigma_{\rm tot}^{\phi}={\pi {\lambda'}^{2}(\mu_R^2) \over 12 S} \sum_{a,b=q,\bar q,g}\,
\int_\tau^1 {dx_1 \over x_1}\, \int_{\tau/x_1}^1{dx_2\over x_2}\,f_a(x_1,\mu_F^2)\,f_b(x_2,\mu_F^2)\,
\Delta_{ab}\left ( \frac{\tau}{x_1\,x_2},m^2_{\phi},\mu_F^2,\mu_R^2 \right ) 
\nonumber\\[2ex]
&&\mbox{with}\quad \tau=\frac{m^2_{\phi}}{S} \quad\,,\quad S=(P_1+P_2)^2\quad
\,,\quad p_\phi^2=m^2_{\phi}\,,
\label{sigphi}
\end{eqnarray}
where $H_1$ and $H_2$ denote the incoming hadrons and $X$ represents an
inclusive hadronic state.
The parton densities denoted by
$f_c(x_i,\mu_F^2)$ ($c=q,\bar q,g$)
depend on the scaling variables $x_i$ ($i=1,2$) through $p_i=x_i P_i$ and 
the mass factorization scale $\mu_F$.  Here $p_i$ ($i=1,2$) are
the momenta of incoming partons namely quarks, antiquarks and gluons. 
The coupling constant $\lambda'$ gets renormalised at the renormalisation scale
$\mu_R$ due to ultraviolet singularities present in the theory.
The factorisation scale is introduced on the right hand side of the
above equation to separate long distant dynamics from the perturbatively 
calculable short distant partonic coefficient 
functions $\Delta_{ab}$.  $\Delta_{ab}$ depends on both $\mu_R$ and $\mu_F$ 
in such a way that the entire scale dependence goes away to all orders
in perturbation theory when convoluted with appropriate parton densities.
This is due to the fact that the observable on the left hand side of the above equation 
is renormalisation group (RG) invariant with respect to both the scales.  This implies
\begin{eqnarray}
&&\mu^2{d \sigma_{\rm tot}^{\phi} \over d \mu^2} =0, \quad \quad \mu=\mu_F,\mu_R \, ,
\\[2ex]
&&\mu_R^ 2{d 
\over d \mu_R^2} \Big[{\lambda'}^2(\mu_R^2) \Delta_{ab}\left (x,m^2_{\phi},\mu_F^2,\mu_R^2\right ) 
\Big]=0\, .
\end{eqnarray}
The partonic coefficient functions that appear in eqn.(\ref{sigphi})
are computable in perturbative QCD in terms of strong coupling constant $g_s$.
The ultraviolet singularities present in the theory are regularised in dimensional
regularisation and are removed in ${\overline {MS}}$ scheme, introducing 
the renormalisation scale $\mu_R$ at every order in perturbative expansion.  
In addition, the Yukawa coupling $\lambda'$ also gets renormalised due to 
strong interaction dynamics.  Hence, for our computation, we require only 
two renormalisation constants 
to obtain UV finite partonic coefficient functions, $\Delta_{ab}$.  
These constants are denoted by
$Z(\mu_R)$ and $Z_{\lambda'}(\mu_R)$, where the former renormalises the strong coupling constant $g_s$ 
and the later Yukawa coupling $\lambda'$
and  
both the couplings $a_s(=g_s/(4\pi))$ (and $\lambda'$) evolve with scale to NNLO through renormalisation group equations:
\begin{eqnarray}
\mu_R^2 {d \over d\mu_R^2} \ln a_s(\mu_R^2) &=&
-\sum_{i=1}^\infty a^i_s(\mu_R^2)~ \beta_{i-1} \,,
\nonumber\\
\mu_R^2 {d \over d\mu_R^2} \ln \lambda'(\mu_R^2)&=&
-\sum_{i=1}^\infty a^i_s(\mu_R^2)~ \gamma_{i-1}\,.
\label{REASLAM}
\end{eqnarray}
where coefficients $\beta_i$ for $i=0,...,3$ 
can be found in \cite{4loop_beta_func}
for $SU(N)$ QCD.
The anomalous dimensions $\gamma_i$ for $i=0,...,3$ can be obtained from
the quark mass anomalous dimensions given in 
\cite{vanRitbergen:1997va}. 
The perturbatively calculable $\Delta_{ab}$ can be expanded in powers of
strong coupling constant $a_s(\mu_R^2)$ as
\begin{eqnarray}
\Delta_{ab}\left (x,m^2_{\phi},\mu_F^2,\mu_R^2\right )=\sum_{i=0}^\infty a_s^i(\mu_R^2)
\Delta_{ab}^{(i)}\left (x,m^2_{\phi},\mu_F^2,\mu_R^2\right )\,.
\nonumber
\end{eqnarray}
$\Delta_{ab}$ gets contributions from various partonic reactions.

The calculation of various contributions from the partonic reactions 
involves careful handling of divergences that result from
one\cite{1loop} and two loop\cite{2loop} integrations in the virtual
 processes  
and two and three body phase space integrations in the real emission processes.
The loop integrals often give ultraviolet, soft and collinear
divergences.  But the phase space integrals give only soft and collinear 
singularities.  Soft divergences arise when the  
momenta of the gluons become zero while the collinear diverges
arise due to the presence of massless partons.   We have regulated all the integrals in
dimensional regularisation with space time dimension $n=4+\ep$.  
The singularities manifest themselves as poles in $\ep$.

We have reduced all the one loop tensorial integrals
to scalar
integrals using the method of Passarino-Veltman \cite{Pas_Velt} in $4+\ep$
 dimensions
and evaluated resultant scalar integrals exactly.  
The $2$-loop form factor, ${\cal F}_\phi(m_\phi^2,\mu^2)$, 
is calculated using the dispersion
technique \cite{Cutkosky:1960sp}.
Two and three body phase space integrals are done
by choosing appropriate Lorentz frames\cite{2.3.Body_phase_space}.
Since we integrate over the total phase space the integrals are Lorentz
invariant and therefore frame independent. 
Several routines are made using the algebraic manipulation program FORM\cite{form} 
in order to perform tensorial reduction of one loop integrals and 
two and three body phase space integrals.

The UV singularities go away after performing renormalisation through
the constants $Z$ and $Z_{\lambda'}$.  
The soft singularities cancel among virtual and real emission processes\cite{BN}
at every order in perturbation theory.  The remaining 
collinear singularities are renormalised systematically using
mass factorisation\cite{KLN}.  
For more details on the computation of NNLO QCD corrections to process of
the kind considered here can be found in \cite{nnlody,prakash}. The full analytical
results for NNLO calculation for sneutrino and/or charge slepton can be found out in our original paper\cite{prakash}.

\section{Results and Discussion}

We considered only the contributions from  
the first generation of quarks.   
Since at hadron colliders, the resonant production is through 
the interaction term $\lambda'_{ijk}L_iQ_jD_k^c$ in the Lagrangian (see eq.(\ref{eq:superpot})), 
for $j,k = 2,3$, the production rate will be suppressed due to the 
low flux of the sea quarks. 
To obtain the production cross section to a particular order,
one has to convolute the partonic coefficient functions $\Delta_{ab}$  with the corresponding 
parton densities $f_a$, both to the same order.   Further 
the coupling constants $a_s(\mu_R)$ and $\lambda'(\mu_R)$ 
should also be evaluated using the corresponding RGEs (eqn.(\ref{REASLAM})) computed to the same order (more details see Ref\cite{vanRitbergen:1997va,Harlander} ). We have used the latest MSTW parton densities \cite{MSTW2008} in our numerical code and the
corresponding values of $\alpha_s(M_Z)$ for LO, NLO and NNLO provided with
the sets. 
\begin{figure}
\vspace{-5cm}
\centering
\includegraphics[scale=0.6]{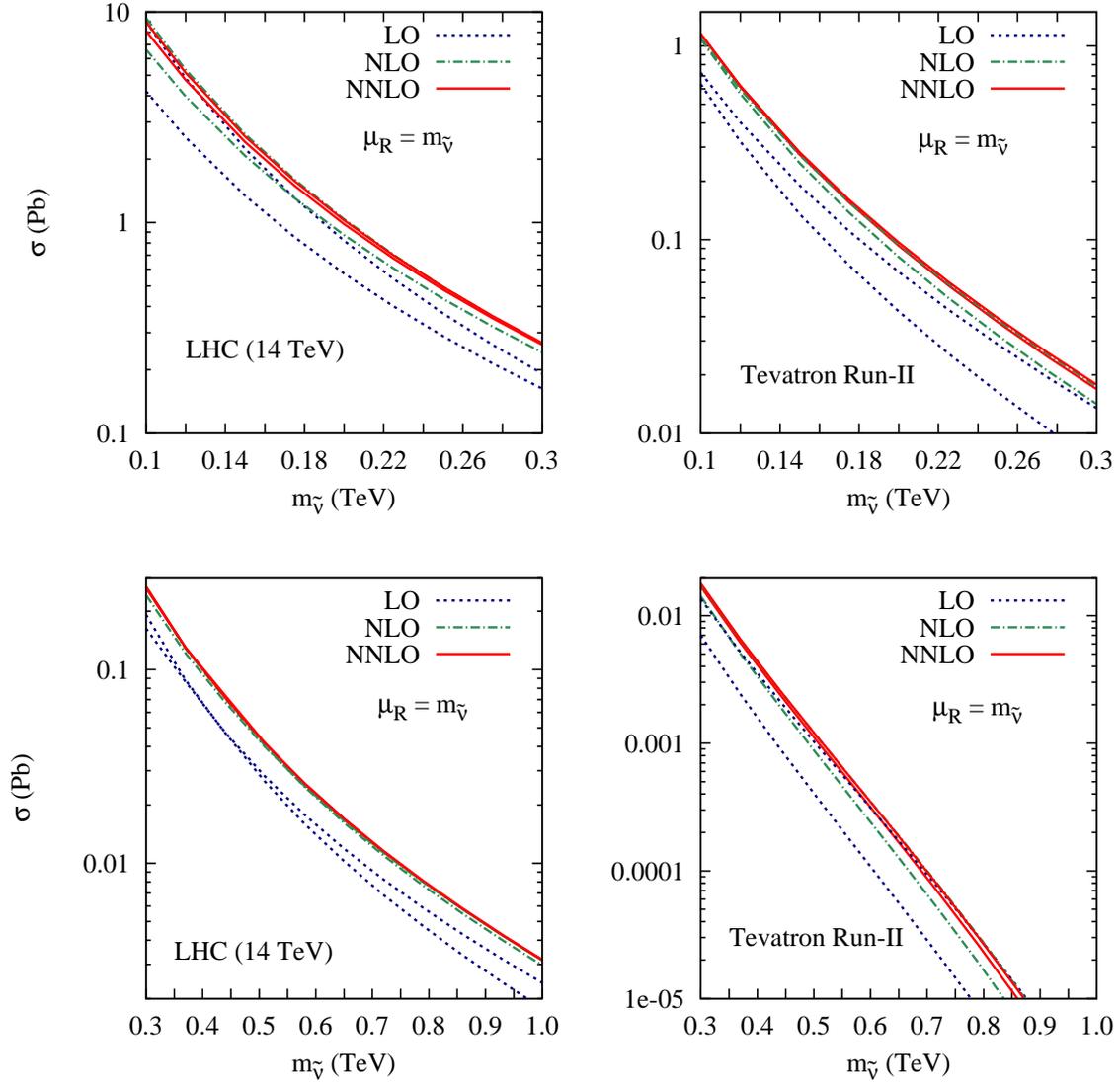}
\vspace{-4cm}
\caption{\em Total cross-section for the ${\tilde \nu}$ production as a function of 
$m_{\tilde{\nu}}$.
For smaller values of sneutrino mass the upper (lower) set of lines 
correspond to the
factorisation scale $\mu_F = 10 m_{\tilde{\nu}}(0.1 m_{\tilde{\nu}})$.  For larger values of sneutrino mass 
the lines cross
each other.
 }
\label{fig:l0}
\end{figure}
Since we are considering one $\lambda'_{i11}$ non-zero, the LO and NLO
cross sections get contributions only from $d \overline d$ , $d g$ and $\overline d g$
initiated subprocesses and no other quark (antiquark) flavors  
contribute to this order.  
At NNLO level, the incoming quarks other than $d$ type quarks
can also contribute.
The total sneutrino production cross section as function of its mass
is plotted in fig.\ \ref{fig:l0} for LHC (left panel) and Run II of Tevatron (right panel) energies.  
We have set the renormalisation scale to be the mass of the
sneutrino, $\mu_{R}=m_{\tilde{\nu}}$.
The pair of lines corresponds to the two extreme choices of factorisation scale: 
$\mu_{F}=10~m_{\tilde{\nu}}$ (upper) and $\mu_{F}= m_{\tilde{\nu}}/10$ (lower). 
The plots clearly demonstrate that the NNLO contributions reduce 
the factorisation scale dependence improving the theoretical predictions
for sneutrino production cross section.

The cross section falls off with the sneutrino
mass due to the availability of phase space with respect to the mass,
the choice of $\mu_R=m_{\tilde{\nu}}$ and the parton densities.  The latter effect, 
understandably, is more pronounced at the Tevatron than at the LHC. 
 
\begin{figure}
\vspace{-4.5cm}
\includegraphics[scale=0.5]{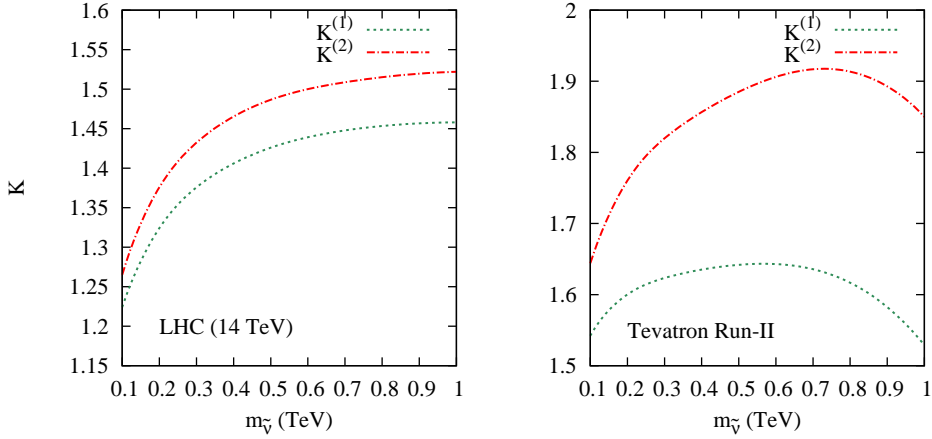}
\vspace{-9cm}
\caption{\em NLO K-factor $K^{(1)}$ and NNLO K-factor $K^{(2)}$ are plotted 
for sneutrino production
at the LHC (left panel) and the Tevatron Run-II (right panel)
as a function of its mass. 
 }
\label{fig:kl0}
\end{figure}

In order to estimate the magnitude of the QCD corrections at NLO and NNLO, we 
define the K-factors as follows: 
$$K^{(1)} = \sigma^\phi_{tot,\rm NLO} / \sigma^\phi_{tot,\rm LO}\, , \hspace{0.5cm} 
K^{(2)} = \sigma^\phi_{tot,\rm NNLO} / \sigma^\phi_{tot,\rm LO}.$$ 
In fig.\ref{fig:kl0}, we have plotted both $K^{(i)}$ ($i=1,2$) as a function
of sneutrino mass.  We have chosen $\mu_F=
\mu_R=m_{\tilde \nu}$ for this study. 
At the LHC, The $K^{(1)}$ varies between $1.23$ to $1.46$ and $K^{(2)}$ between
$1.27$ to $1.52$ in the mass range $100~GeV \le m_{\title \nu} \le 1~TeV$.  
At the Tevatron, we find that $K^{(1)}$ varies between $1.55$ to $1.53$ and $K^{(2)}$ between
$1.65$ to $1.85$ for the same mass range.
Note that numbers for $K^{(1)}$ differ from those given in our earlier work \cite{swapan}
due to the running of $\lambda'$ in the present analysis.  
The present analysis using running $\lambda'$ is the correct way to 
reduce renormalisation scale dependence in the cross section.    
We also observe that $K$ factor is much bigger at the Tevatron compared to that of at the LHC. 
The reason behind this is attributed to the different behavior of parton densities at the Tevatron and the LHC.
Note that parton densities rise steeply as $x \rightarrow 0$ and fall off very fast as $x \rightarrow 1$, 
which means the dominant contribution to the production results from the phase space region where 
$x \sim \tau (= m^2_{\tilde \nu}/S)$ becomes small.
$\tau$ at Tevatron ($0.05\lsim\tau \lsim 0.5$) is larger compared to
that at LHC ( $0.007\lsim\tau\lsim 0.07$) (see also fig.\ref{fig:kl0}). 
Because of this, at Tevatron the valence quark initiated processes dominate while 
gluon and sea quark initiated processes dominate at the LHC. 
As the mass of the sneutrino increases, that is $x$ approaches to unity,
the $K$-factor at Tevatron naturally falls off. 
At LHC, in the higher mass region ($\sim 1$ TeV),
valence quark densities start to dominate and hence it stays 
almost flat compared to Tevatron. 
\begin{figure}
\vspace{-4.5cm}
\includegraphics[scale=0.5]{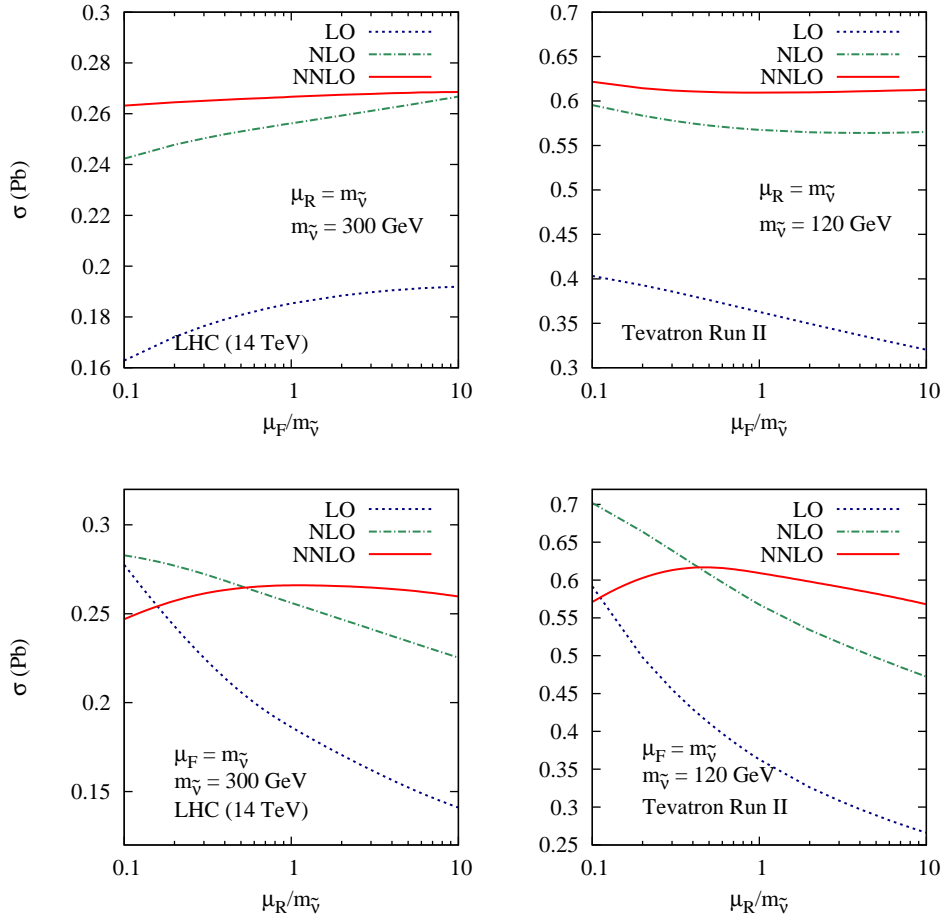}
\vspace{-2.5cm}
\caption{\em In the upper panel, sneutrino production cross sections are plotted against the 
factorisation scale $\mu_F$ with a fixed renormalisation scale $\mu_R=m_{\tilde \nu}$ 
for both LHC and Tevatron energies.  In the lower panel,  
they are plotted against the 
renormalisation scale $\mu_R$ with a fixed factorisation scale $\mu_F=m_{\tilde \nu}$ 
for both LHC and Tevatron energies.   The mass of the sneutrino is taken to be $300$ GeV ($120$ GeV) at LHC
 (Tevatron).
 }
\label{fig:ml0}
\end{figure}
We now turn to study the impact of   
the factorisation scale $(\mu_F)$ and 
the renormalisation scale ($\mu_R$) on the production cross section. 
The factorisation scale dependence for both LHC (left panel) 
and Tevatron (right panel) are shown in upper panels of fig.\ \ref{fig:ml0}, 
for $m_{\tilde \nu}=300~GeV$ (LHC), $m_{\tilde \nu}=120~GeV$ (Tevatron).  We have 
chosen $\mu_R=m_{\tilde \nu}$ for both the LHC and the Tevatron.   
The factorisation scale 
is varied between $\mu_F=0.1 ~m_{\tilde \nu}$ and $\mu_F=10 ~m_{\tilde \nu}$.  
We find that the factorisation scale dependence decreases 
in going from LO to NLO to NNLO as expected.   

The dependence of the renormalisation scale dependence on the total
cross sections for the resonant production of sneutrino at the LHC and the Tevatron
is shown in the lower panels of fig.\ \ref{fig:ml0}.
Note that the LO is already $\mu_R$ dependent  
due to the coupling $\lambda' (\mu_R)$. 
We have performed this analysis for sneutrino mass
$m_{\tilde{\nu}}\,
= 300~GeV$ (LHC), $m_{\tilde \nu}=120~GeV$ (Tevatron).  
We have set the factorisation scale $\mu_F=m_{\tilde \nu}$ 
and
the renormalisation scale is varied in the  
range $0.1 \le \mu_R/m_{\tilde \nu} \le 10$. 
We find significant 
reduction in the $\mu_R$ scale dependence when higher
order QCD corrections are included. 
It is clear from both the panels of fig.\ \ref{fig:ml0}  that
our present NNLO result makes the predictions almost independent
of both factorisation and renormalisation scales.

    We could not discuss or show the results of charged slepton due to
page limitation. We request reader to follow
the Ref.\cite{prakash}.

{\bf Acknowledgement:} Speaker acknowledges the full support of Saha
Institute of Nuclear Physics, India where this work has been done. Speaker
also thanked RADCOR's organiser for inviting him to give a seminar in RADCOR 2011.

\newcommand{\prep}[3]{Phys.Rept. {\bf #1} (#3) #2}
\newcommand{\np}[3]{Nucl.Phys. {\bf B#1} (#3) #2}                     




\begin{thebibliography}{99}



\bibitem{SUSY} H.P.~Nilles, \prep{110}{1}{1989};
  H.E.~Haber and G.L.~Kane, \prep{117}{75}{1985};
  S.~Dawson, \np{261}{297}{1985}.

\bibitem{ATLAS} ATLAS Colaboration () Eur.Phys.J.{\bf C71} (2011) 1809
 
\bibitem{CDF_col}CDF Collaboration (Darin E. Acosta et al.) Phys.Rev.Lett. 
{\bf 91} (2003) 171602; Phys.Rev.Lett. {\bf 95} (2005) 131801;
CDF Collaboration (A. Abulencia et al.)  Phys.Rev.Lett. 
{\bf 95} (2005) 252001; Phys.Rev.Lett. {\bf 96} (2006) 211802;
 CDF Collaboration (T. Aaltonen et al.)  Phys.Rev.Lett. {\bf 102}
(2009) 091805.
 
\bibitem{D0_col} D0 Collaboration (V.M. Abazov et al.) 
Phys.Rev.Lett. {\bf 97} (2006) 111801. 


\bibitem{swapan} Debajyoti Choudhury, Swapan Majhi and V. Ravindran
 Nucl.Phys. {\bf B660} (2003) 343; JHEP {\bf 0601} (2006) 027.

\bibitem{4loop_beta_func} S.A. Larin, J.A.M. Vermaseren, Phys.Lett. {\bf B303} 
                        (1993) 334;
                    T. van Ritbergen, J.A.M. Vermaseren, S.A. Larin,
                      Phys.Lett. {\bf B400} (1997) 379;
  M.~Czakon, Nucl.Phys. {\bf B710} (2005) 485.

\bibitem{vanRitbergen:1997va} 
  S.~P.~Martin and M.~T.~Vaughn, Phys.\ Rev.\  D {\bf 50} (1994) 2282
  [Erratum-ibid.\  D {\bf 78} (2008) 039903];
J.A.M. Vermaseren, S.A. Larin, T. van Ritbergen,
                   Phys.Lett. {\bf B405} (1997) 327; 
  B.~C.~Allanach, A.~Dedes and H.~K.~Dreiner,
  Phys.\ Rev.\  D {\bf 60} (1999) 056002.



\bibitem{1loop}
    T. Matsuura, thesis Leiden University, 1989;
    S. Dawson, Nucl.Phys. {\bf B359} (1991) 283;
    A. Djouadi, M. Spira, P. Zerwas, Phys.Lett. {\bf B264} (1991) 440.


\bibitem{2loop}
    G. Kramer and B. Lampe, Z.Phys. {\bf C34} (1987) 497
    [Erratum: {\bf C42} (1989) 504];
    T. Matsuura and W.L. van Neerven,  Z.Phys. {\bf C38} (1988) 623;
    T. Matsuura S.C. van der Marck and W.L. van Neerven, Phys.Lett.
    {\bf B211} (1988) 171;  Nucl.Phys. {\bf B319} (1989) 570;
    W.L. van Neerven,  Nucl.Phys. {\bf B268} (1986) 453;
    R.J. Gonsalves,  Phys.Rev. {\bf D28} (1983) 1542 ;
    R.V. Harlander, Phys.Lett. {\bf B492} (2000) 74.

\bibitem{Pas_Velt} G. Passarino and M. J. G. Veltman, 
                   Nucl.Phys. {\bf B160} (1979) 151

\bibitem{Cutkosky:1960sp}
  R.~E.~Cutkosky, J.\ Math.\ Phys.\  {\bf 1} (1960) 429;
  W.~L.~van Neerven, Nucl.\ Phys.\  B {\bf 268} (1986) 453;
  V.~Ravindran, J.~Smith and W.~L.~van Neerven,
  Nucl.\ Phys.\  B {\bf 704} (2005) 332.


\bibitem{2.3.Body_phase_space}
    T. Matsuura and W.L. van Neerven, Z.Phys. {\bf C38} (1988) 623;
    T. Matsuura S.C. van der Marck and W.L. van Neerven, Phys.Lett.
    {\bf B211} (1988) 171;  Nucl.Phys. {\bf B319} (1989) 570;
    T. Matsuura, thesis Leiden University, 1989;
    R.K.Ellis, M.A.Furman, H.E. Haber and I. Hinchliff, Nucl.Phys. 
     {\bf B173} (1980) 397;
    J. Smith, D. Thomas and W.L. van Neerven,  Z.Phys. {\bf C44} (1989) 267;
    W. Beenakker, H. Kuijf, W.L. van Neerven, J. Smith, Phys.Rev. {\bf D40}
     (1989) 54;
    V. Ravindran, J. Smith, W.L. van Neerven, Pramana {\bf 62} (2004) 683.

\bibitem{form}
FORM by J.A.M. Vermaseren, version 3.0 available from
http://www.nikhef.nl/form;
  arXiv:math-ph/0010025.


\bibitem{BN} F. Block and A. Nordsieck,  Phys.Rev. {\bf 52} (1937) 54;
            D.R. Yannie, S.C. Frautschi and H. Suura, 
             Ann.Phys.(N.Y.) {\bf 13} (1961) 379.

\bibitem{KLN} T. Kinoshita,  J.Math.Phys. {\bf 3} (1962) 650;
            T.D. Lee and M. Nauenberg,  Phys.Rev. {\bf 133} (1964) B1549;
            N. Nakanishi,  Prog.Theor.Phys. {\bf 19} (1958) 159

\bibitem{nnlody}
  R.~Hamberg, W.~L.~van Neerven and T.~Matsuura,
  Nucl.\ Phys.\  B {\bf 359} (1991) 343;
  R.~V.~Harlander and W.~B.~Kilgore,
  Phys.\ Rev.\ Lett.\  {\bf 88} (2002) 201801;
  C.~Anastasiou and K.~Melnikov,
  Nucl.\ Phys.\  B {\bf 646} (2002) 220;
  V.~Ravindran, J.~Smith and W.~L.~van Neerven,
  Nucl.\ Phys.\  B {\bf 665} (2003) 325.
 
\bibitem{prakash} Swapan Majhi, Prakash Mathews and V. Ravindran,
Nucl.Phys.{\bf B850} (2011) 287. 

\bibitem{Harlander} Robert V.~Harlander and William B.~Kilgore
Phys.Rev. {\bf D68} (2003) 013001.

\bibitem{MSTW2008} A.D. Martin, W.J. Stirling, R.S. Thorne and G. Watt
Eur.Phys.J. {\bf C63} (2009) 189,
Eur.Phys.J. {\bf C64} (2009) 653.


\end{thebibliography}
\end{document}